%% file: susceptometer_arxiv.tex
\documentclass[
 aip,
 rsi,%
 amsmath,amssymb,
preprint,
superscriptaddress,
floatfix,
]{revtex4-1}

\usepackage{graphicx}
\usepackage{dcolumn}
\usepackage{bm}

\begin{document}

\preprint{AIP/123-QED}

\title[SQUID susceptometers]{Scanning SQUID susceptometers with sub-micron spatial resolution}
\author{John R. Kirtley}%
 \email{jkirtley@stanford.edu.}
\affiliation{Dept. of Applied Physics, Stanford University, Stanford, California 94305-4045, USA}
\author{Lisa Paulius}
 \affiliation{Dept. of Physics, Western Michigan University, Kalamazoo, Michigan 49008-5252, USA}
\author{Aaron J. Rosenberg} 
\affiliation{Dept. of Applied Physics, Stanford University, Stanford, California 94305-4045, USA}
\author{Johanna C. Palmstrom} 
\affiliation{Dept. of Applied Physics, Stanford University, Stanford, California 94305-4045, USA}
\author{Connor M. Holland} 
\affiliation{Dept. of Applied Physics, Stanford University, Stanford, California 94305-4045, USA}
\author{Eric M. Spanton}
\affiliation{Dept. of Physics, Stanford University, Stanford, California 94305-4045, USA}
\author{Daniel Schiessl}
\affiliation{Attocube Systems AG,
K{\"o}niginstra{\ss}e 11a, 80539 Munich, Germany}
\author{Colin L. Jermain} 
\affiliation{Dept. of Physics, Cornell University, Cornell, Ithaca, New York  14853, USA}
\author{Jonathan Gibbons} 
\affiliation{Dept. of Physics, Cornell University, Cornell, Ithaca, New York  14853, USA}

\author{Y.-K.-K. Fung}
\affiliation{IBM Research Division, T.J. Watson Research Center, Yorktown Heights, New York 10598, USA}
\author{Martin E. Huber} 
\affiliation{Dept. of Physics, University of Colorado Denver, Denver, Colorado 80217-3364, USA}
\author{Daniel C. Ralph} 
\affiliation{Dept. of Physics, Cornell University, Cornell, Ithaca, New York  14853, USA}
\affiliation{Kavli Institute at Cornell, Ithaca, New York, 14853, USA}
\author{Mark B. Ketchen}
\affiliation{OcteVue, Hadley, Massachusetts 01035, USA}
\author{Gerald W. Gibson, Jr.}
\affiliation{IBM Research Division, T.J. Watson Research Center, Yorktown Heights, New York 10598, USA}
\author{Kathryn A. Moler}
\affiliation{Dept. of Applied Physics, Stanford University, Stanford, California 94305-4045, USA}

\homepage{http://kirtleyscientific.com}

\date{\today}

\begin{abstract}
Superconducting QUantum Interference Device (SQUID) microscopy has excellent magnetic field sensitivity, but suffers from modest spatial resolution when compared with other scanning probes. This spatial resolution is determined by both the size of the field sensitive area and the spacing between this area and the sample surface. In this paper we describe scanning SQUID susceptometers that achieve sub-micron spatial resolution while retaining a white noise floor flux sensitivity of $\approx 2\mu\Phi_0/Hz^{1/2}$. 
This high spatial resolution is accomplished by deep sub-micron feature sizes, well shielded pickup loops fabricated using a planarized process, and a deep etch step that minimizes the spacing between the sample surface and the SQUID pickup loop. We describe the design, modeling, fabrication, and testing of these sensors. Although sub-micron spatial resolution has been achieved previously in scanning SQUID sensors, our sensors not only achieve high spatial resolution, but also have integrated modulation coils for flux feedback, integrated field coils for susceptibility measurements, and batch processing. They are therefore a generally applicable tool for imaging sample magnetization, currents, and susceptibilities with higher spatial resolution than previous susceptometers.
\end{abstract}

\maketitle

\section{\label{introduction}Introduction}
A SQUID is a superconducting ring interrupted by one or two Josephson weak links.
SQUID microscopy\cite{rogers1983deo,vu1993dis,black1993mmu,kirtley1995hrs, kirtley1999ssm,kirtley2010fss} scans a SQUID to image the magnetic flux above sample surfaces. It has the advantages of high sensitivity, an easily calibrated, linear response to magnetic flux, and minimal interaction between the sensor and the sample.  The spatial resolution of a scanning SQUID magnetometer is determined by the area of either the SQUID itself or of a pickup loop integrated into the SQUID, as well as the spacing between this area and the sample surface. The sensitivity of a SQUID magnetometer to a localized field source is also determined in part by these two factors.

There are two principle strategies for achieving high spatial resolution in SQUID microscopy. The first is to use small SQUIDs.\cite{veauvy2002smus,granata2016nano} Such small SQUIDs are made either by fabricating  them from a single planar superconducting layer, with the Josephson junctions composed of narrow constrictions,\cite{wernsdorfer1997een,troeman2007nbn,hao2008man} or by using the ``SQUID on a tip" technology, in which a SQUID is fabricated by evaporating superconductors onto the end of a hollow pulled glass cylinder.\cite{finkler2010san,vasyukov2013scanning} This strategy has the advantages of simplicity and no spacing layers between the flux sensing area and the sample surface, but the disadvantages that 1) a feedback flux to the SQUID to operate at the most sensitive flux position and linearize the response would also apply a large field to the sample itself, and 2) to date these sensors do not make local susceptibility measurements.

The second strategy for producing high spatial resolution scanning SQUIDs, which is the one that we follow in the present work, is to integrate a small pickup loop into a more conventional SQUID.\cite{vu1993dis,tsuei1994psf,ketchen1995dap,kirtley1995hrs,koshnick2008ats} This strategy allows incorporation of both a flux modulation coil into the body of the SQUID and a field coil near the pickup loop for making local susceptibility measurements.\cite{ketchen1984mss,gardner2001ssq} In this paper we describe SQUID susceptometers that achieve full-widths at half-maximum of 0.5 $\mu$m in images of magnetic nanoparticles. This high spatial resolution is accomplished by deep sub-micron feature sizes, well shielded pickup loops fabricated using a planarized process, and a deep etch step that makes it possible for the surface of the SQUID sensor directly above the pickup loop to contact the sample surface. We describe the design, modeling,  fabrication, and testing of these sensors.

\section{\label{design}Layout and design}
\subsection{Layout}
\begin{figure}
\includegraphics[width=0.5\textwidth]{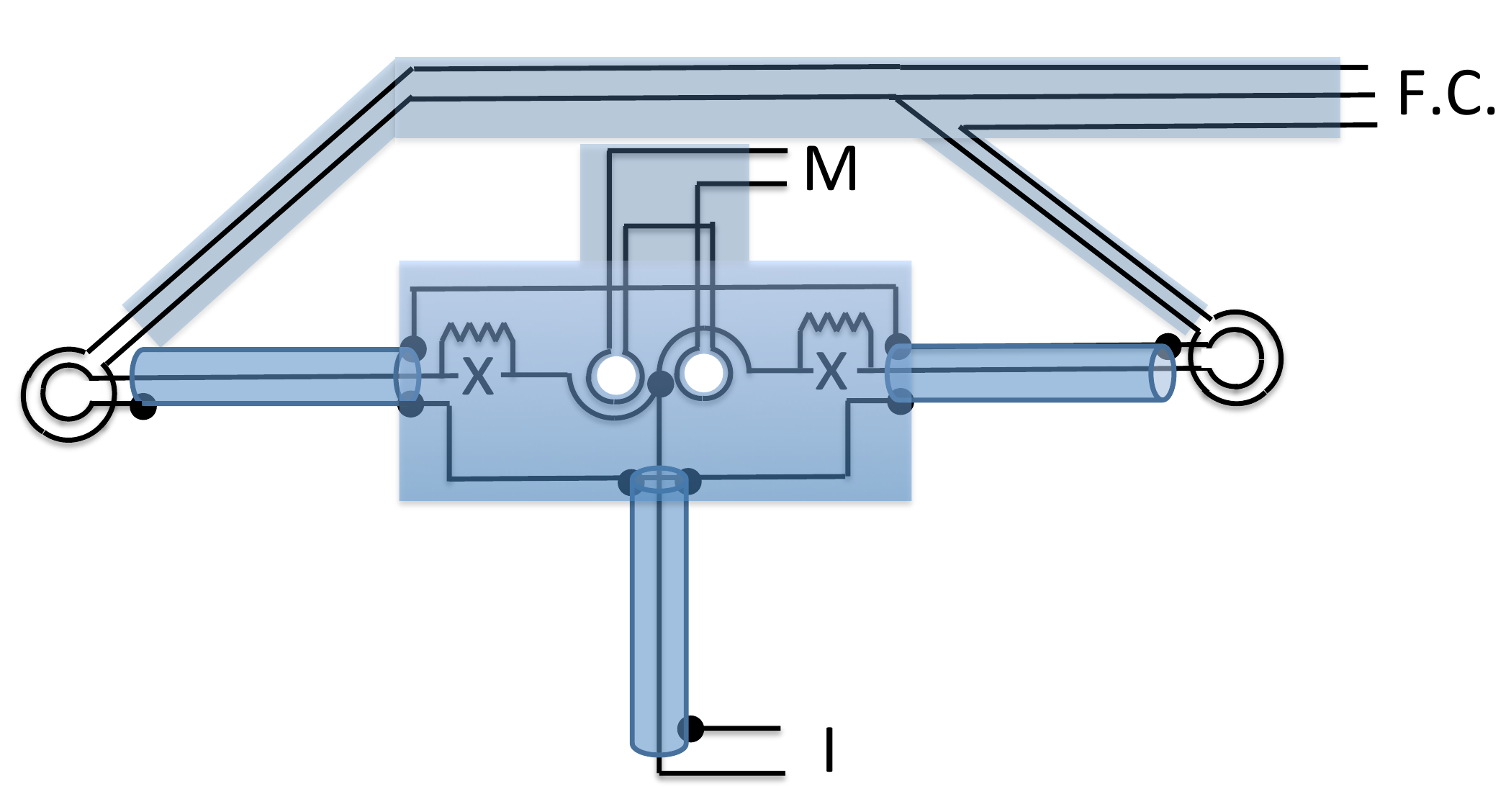}

\caption{Schematic for our susceptometers, showing the bias current leads ($I$), the modulation coil leads ($M$), the field coil leads ($F.C.$), and the Josephson junctions ($X$). The semi-transparent regions represent superconducting shielding. The central junction/shunt/modulation coil region, as well as the current leads and the leads between the pickup loops and the junction region are shielded with superconducting coaxes. The field coil and modulation coil leads are on top of shielding ground planes.
}
\label{fig:schematic}
\end{figure}

A schematic of our susceptometers is shown in Figure \ref{fig:schematic}. The basic layout follows closely that of Huber et al. \cite{huber2008gms}, which was in turn based on previous susceptometer designs.\cite{ketchen1989dfa, gardner2001ssq}  This layout has a gradiometric design, such that the resultant SQUIDs are insensitive to uniform magnetic fields. The modulation coils are integrated into the body of the SQUID, and single turn field coils surrounding each pickup loop apply magnetic fields to the sample for local, gradiometric susceptibility measurements. This layout has low sensitivity to uniform fields and small parasitic inductance in series with the junctions, although there is high parasitic capacitance in parallel with the junctions. 

\begin{figure}
\includegraphics[width=0.5\textwidth]{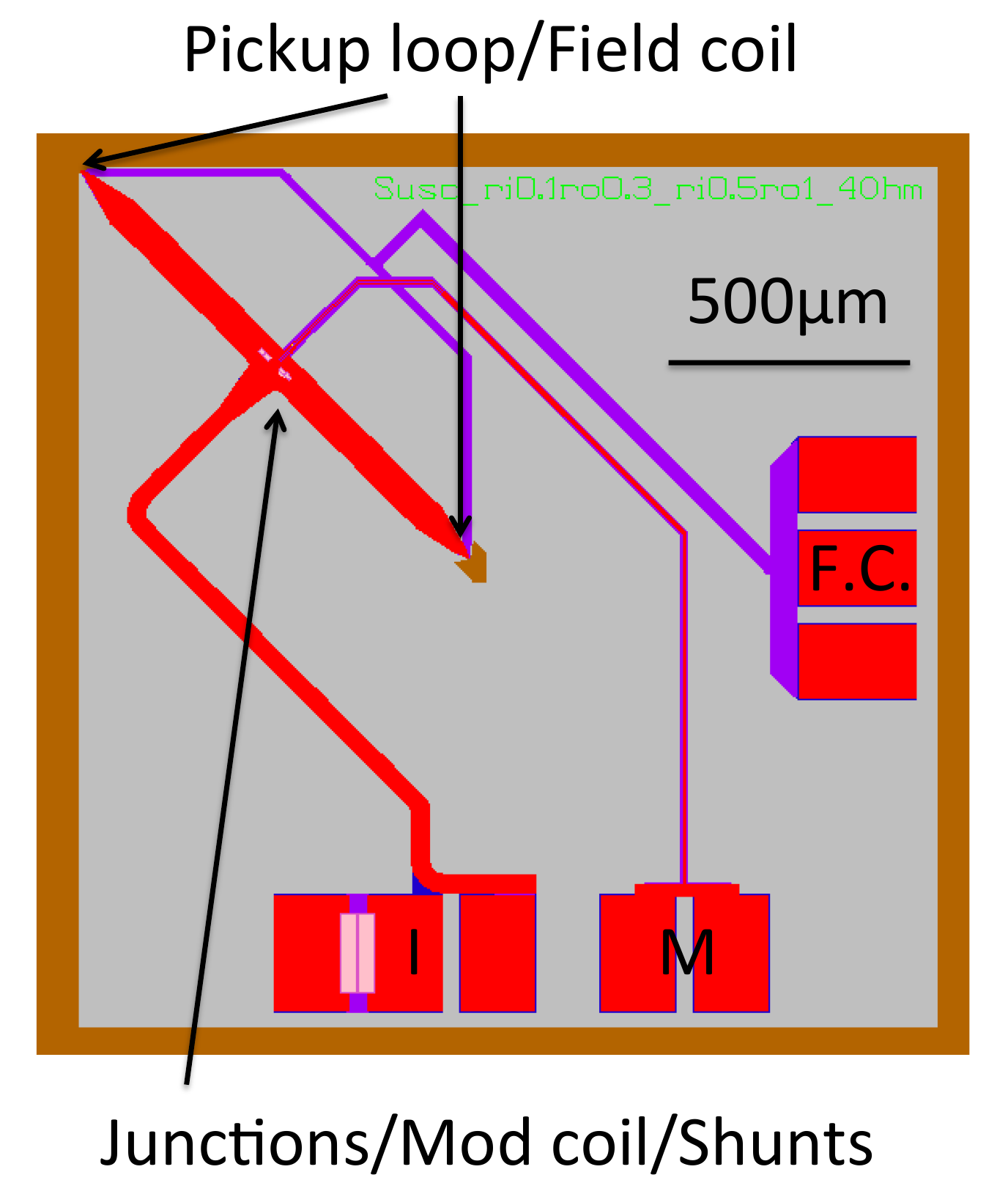}

\caption{ Full chip layout of our susceptometers. The current bias, modulation coil, and field coil bonding pads are labelled $I$, $M$, and $F.C.$ respectively. Expanded views of the junction/mod coil/shunt region are in Fig. \ref{fig:center_layouts}, and expanded views of one of the pickup/field coil regions are in Fig. \ref{fig:pickup_layouts}.
}
\label{fig:layouts}
\end{figure}

\begin{figure}
\includegraphics[width=0.8\textwidth]{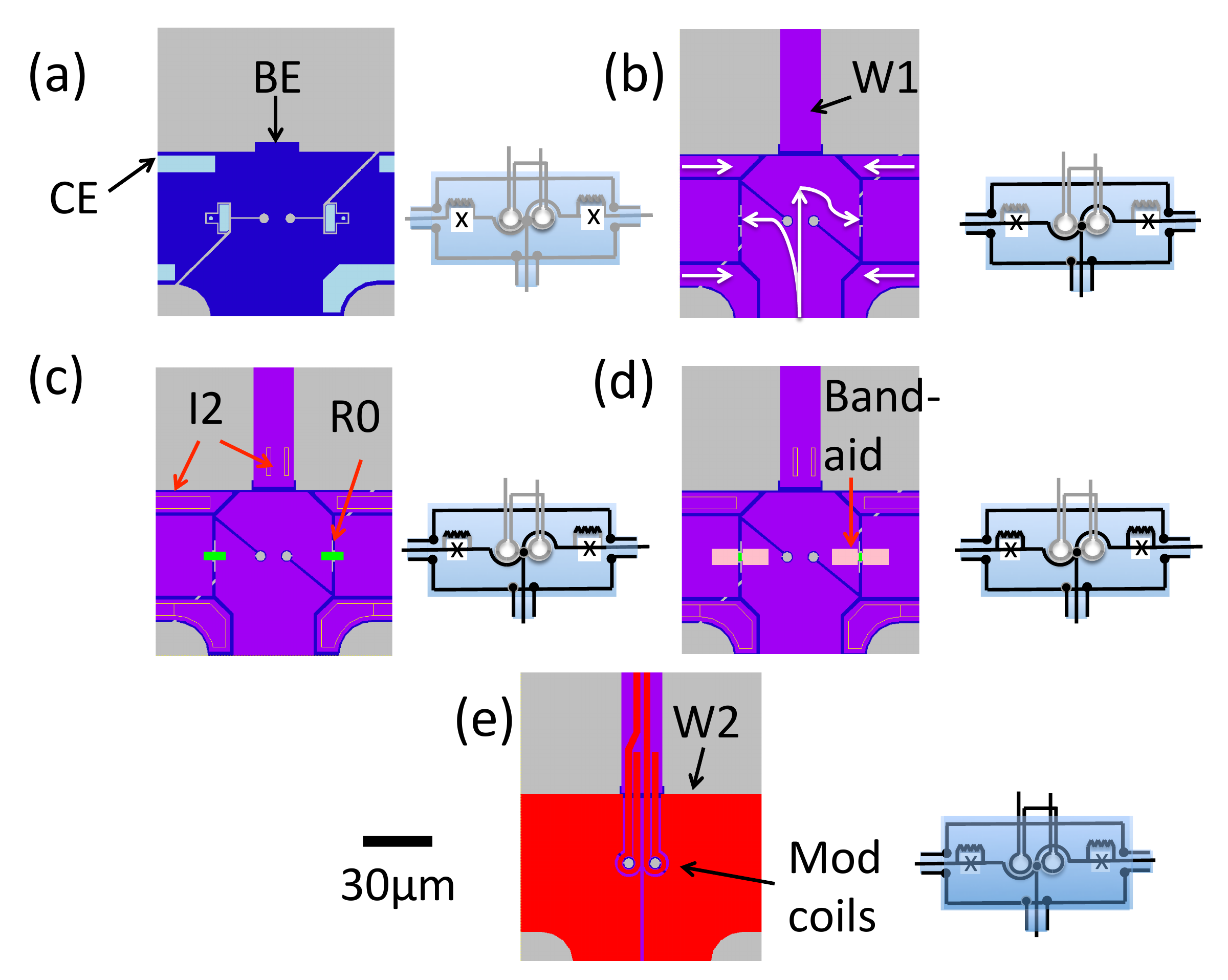}

\caption{ Layout of the central region of our susceptometers, with the panels arranged in order of deposition. Next to each panel is a schematic, with the completed sections black, the uncompleted sections gray, and superconducting shielding in transparent blue. (a) The base electrode ($BE$) and counter-electrode ($CE$) of the tri-layer form the lower shielding layer, the junctions, which are the small dots, and vias to the first wiring level. (b) The first wiring level ($W1$) provides current paths out towards the pickup loops and back as indicated by the arrows. (c) The Au/Pd (R0)  layer forms the shunt resistors, and $I2$ provides vias through the SiO$_2$ between $W1$ and $W2$. (d) The $Band-aid$s connect $W1$ and $R0$.  (e) The second wiring level ($W2$) forms the upper shield and the modulation coils.
}
\label{fig:center_layouts}
\end{figure}

\begin{figure}
\includegraphics[width=0.5\textwidth]{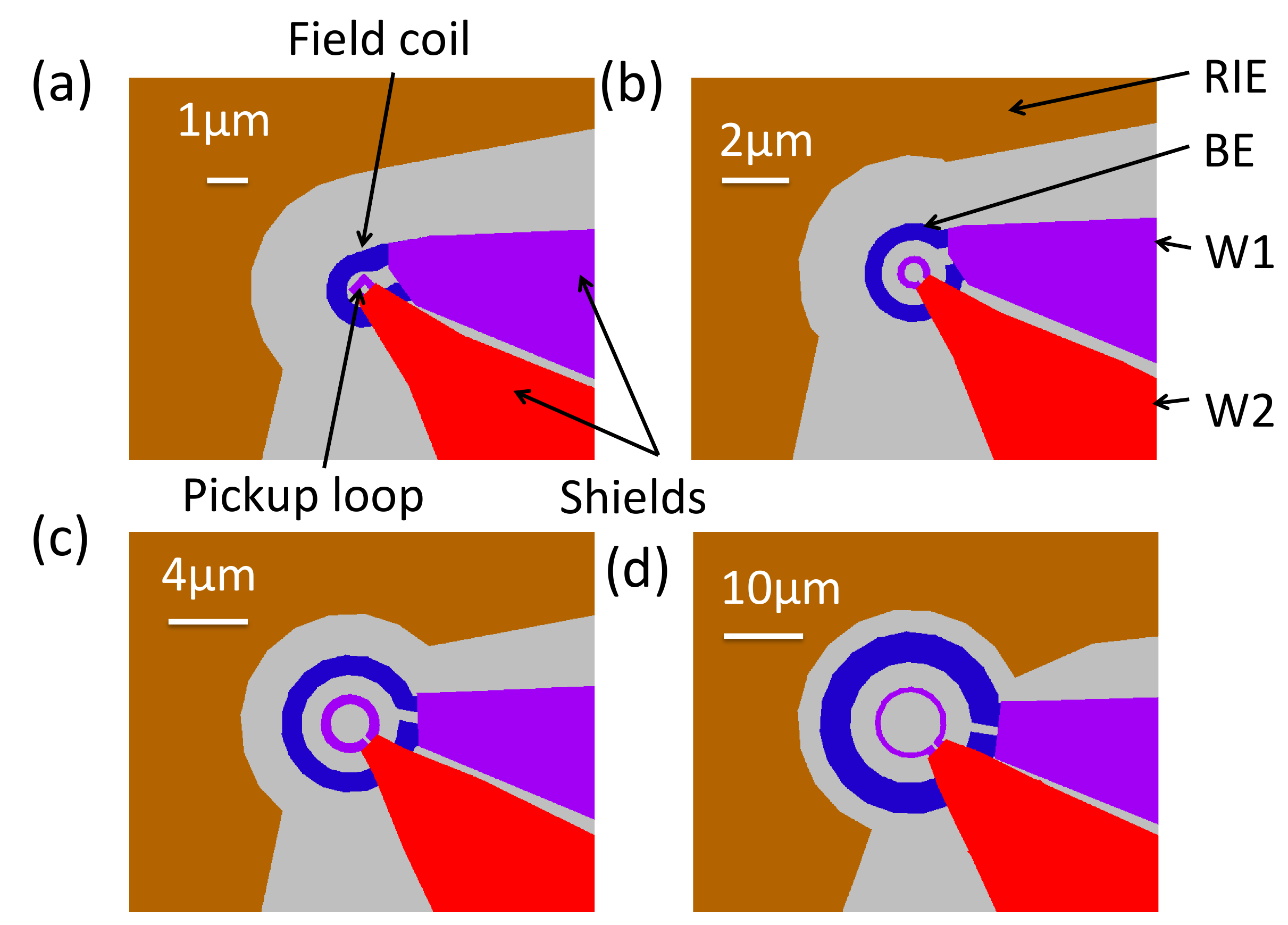}

\caption{ Pickup loop/field coil layouts for 4 different pickup loop sizes. (a) 0.2 $\mu$m inside diameter, (b) 0.6 $\mu$m inside diameter, (c) 2$\mu$m inside diameter, and (d) 6 $\mu$m inside diameter. Selected dimensions are given in Table \ref{table:mutuals}.
}
\label{fig:pickup_layouts}
\end{figure}

\begin{figure}
\includegraphics[width=0.5\textwidth]{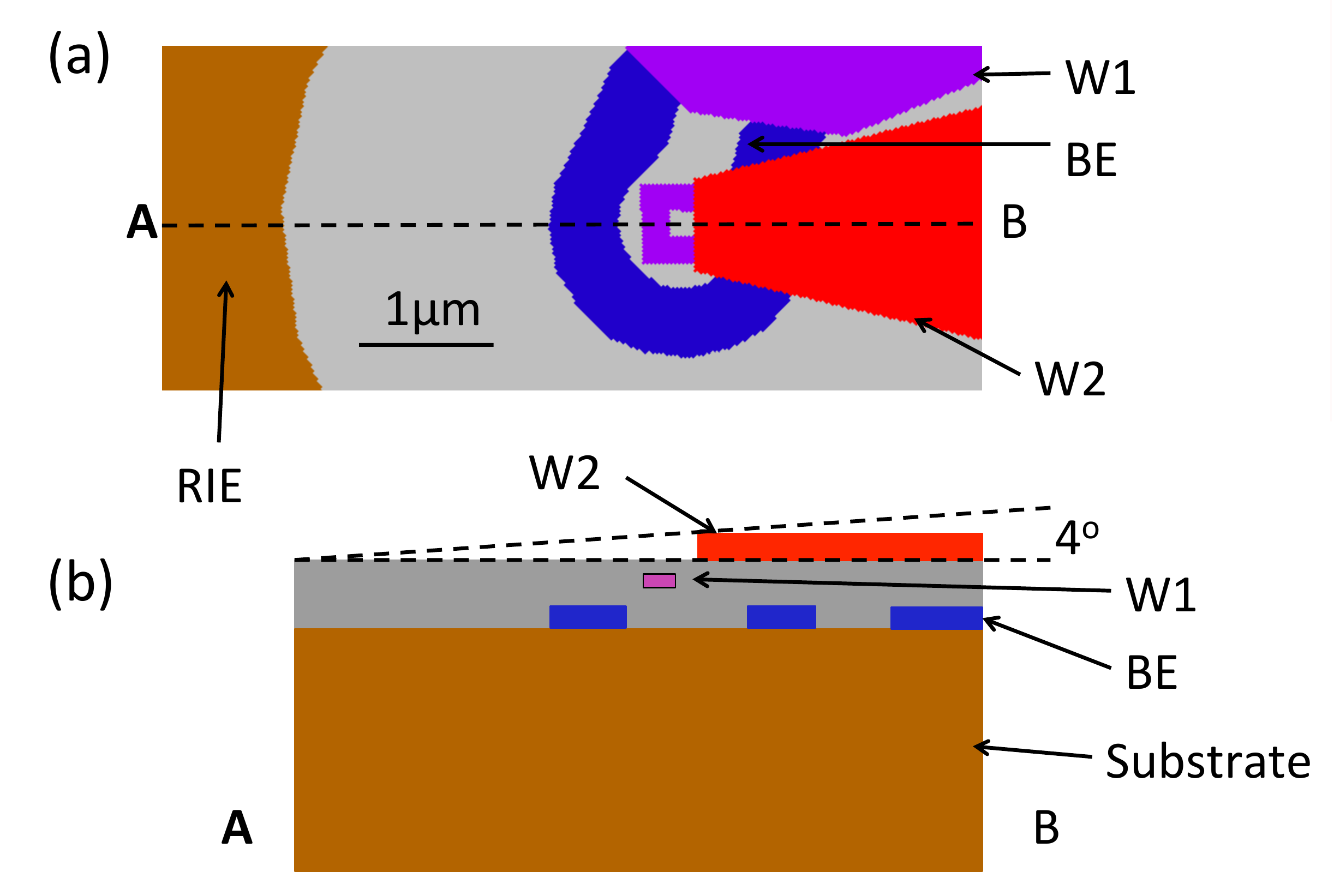}

\caption{ (a) Top view of the layout of our 0.2$\mu$m inside diameter pickup loop susceptometer. (b) Cross-sectional view through the plane labelled by $A-B$ in a). When the angle between the susceptometer and the sample surface is more than 4$^o$ the edge of the 10 $\mu$m deep etch touches the sample first. When the angle is less than 4$^o$ the edge of the $W2$ shield touches first. The minimum spacing between the top of the pickup loop and the sample surface is $\approx$ 0.33 $\mu$m for this susceptometer. 
}
\label{fig:cross_section}
\end{figure}

The layout for our susceptometers is shown in Figure \ref{fig:layouts}. Each chip is 2 mm $\times$ 2 mm in size. The same color scheme is used for the layers in Fig.'s \ref{fig:layouts}, \ref{fig:center_layouts}, \ref{fig:pickup_layouts}, \ref{fig:cross_section}, \ref{fig:three_level_process} and \ref{fig:susceptibility_data}. The layout for the susceptometers is rotated with respect to the earlier  design \cite{huber2008gms} such that one of the pickup loops is close to a corner of the chip. This, combined with a deep etch step, allows close proximity of the pickup loop to the sample surface without further processing. The 500 $\mu$m long leads from the junction area to the pickup loops are superconducting coaxial until the last 50 $\mu$m. The current leads to the SQUID are superconducting coaxes, and the leads to the modulation coil and the field coil are shielded by superconducting ground planes.

An expanded view of the central region of the susceptometer is shown in Fig. \ref{fig:center_layouts}. Fabrication (see Sec. \ref{process}) begins by defining the Nb/Al$_2$O$_3$/Nb trilayer base electrode (BE) and counter-electrode (CE) (Fig. \ref{fig:center_layouts} (a)). In the junction regions are large area trilayer counter-electrodes acting as vias to the base electrode in series with the smaller area junctions. The first wiring level ($W1$)  carries current from the bonding pads, around the modulation coils, through the junctions and shunt resistors out to the pickup loops, and back as indicated by the white arrows in Fig. \ref{fig:center_layouts}(b). Vias through the SiO$_2$ layer (I2, Fig. \ref{fig:center_layouts}(c)) make contact between $W1$ and $W2$ to form coaxial shielding for the pickup loop leads. The Au/Pd resistor layer ($R0$, Fig. \ref{fig:center_layouts}(c)) forms shunt resistors in parallel with the junctions. ``Band-aids" were added during the processing run, when it was discovered that there was poor conductance between $W1$ and $R0$ from underneath, to make low resistance contacts from the top (Fig. \ref{fig:center_layouts}(d)). The second wiring level $W2$ (Fig. \ref{fig:center_layouts}(e)) acts to shield the pickup loop leads. $W2$ also forms the modulation coils, which surround holes in $BE$ that are 5 $\mu$m in diameter, smaller than the 10$\mu$m diameter modulation holes in the earlier design,\cite{huber2008gms} to reduce the total inductance.\cite{koshnick2010dcf}  

In SQUID microscopy there are tradeoffs between spatial resolution and sensitivity that depend in detail on the type of field source.\cite{kirtley2010fss} Therefore we designed and fabricated 
four different pickup loop/field coil pairs (Figure \ref{fig:pickup_layouts}). We will concentrate in this paper on results from the 0.2 $\mu$m inside diameter pickup loop devices shown in Fig. \ref{fig:pickup_layouts} (a). In Figure \ref{fig:pickup_layouts} RIE (reactive ion etch) is the 10$\mu$m deep etch, the field coil is composed of the base electrode ($BE$), the pickup loop and the shield for the field coil are composed of the first wiring level ($W1$), and the upper shield for the pickup loop is composed of the second wiring level $W2$. Not visible is a lower shield composed of $BE$ for the pickup loop. The 0.2 $\mu$m inside diameter pickup loop is the smallest that can be fabricated given the constraints of 0.2 $\mu$m linewidths and spacings in our lithography.

As mentioned above, the spatial resolution in SQUID microscopy is not only set by the size of the effective pickup loop area, but also by the spacing between this area and the field source. In our SQUID susceptometers, this spacing is the sum of two distances (see Figure \ref{fig:cross_section}): The first is the spacing between the top layer of the sensor and the pickup loop. The second is the spacing between the sample surface and the top surface of the sensor. In optimizing our geometry there is a trade-off between minimizing the first spacing, which requires a thin top pickup loop shield ($W2$), and minimizing the pickup from the leads, which requires a thick $W2$ layer. We opted for a $W2$ thickness of 0.2$\mu$m, about twice the London penetration depth of Nb.

The spacing between the 10 $\mu$m deep etch and the center of the 0.2 $\mu$m inside diameter pickup loop is 3 $\mu$m. In addition, the spacing between the 10$\mu$m deep etch and the dicing channel of the chip is about 120 $\mu$m. This means that without additional processing the angle between the susceptometer and the sample must be less than 5$^o$ for the etched edge to touch first. If this angle is less than 4$^o$, the top edge of the W2 shield touches first. At any angle less than 4$^o$, when the sensor is touching the sample the spacing between the sample surface and the top of the pickup loop is approximately 0.33  $\mu$m, which is the sum of the W2 and I2 thicknesses. All of the imaging data presented in this paper was taken with a 0.2 $\mu$m inside diameter pickup loop susceptometer with a deep etch, oriented relative to the sample at an angle less than 4$^o$.

\subsection{Design parameters}
We used the rules established by Tesche and Clark \cite{tesche1977dcs} to choose our susceptometer parameters to minimize noise. We estimate\cite{chang1981nci,whiteley_induct} the inductance of our devices  as follows: the modulation coil region contributes about 35pH, the coaxial leads to the pickup loops about 5.6pH/mm x 0.8mm = 4.5 pH, the transition to the pickup loops about 10pH, and a typical 2 $\mu$m diameter pickup loop another 1.5pH \cite{jaycox1981pcs} - totaling about 50pH. This would imply an optimal critical current of $I_0 = \Phi_0/2 L = 20\mu$A. We targeted our junction critical currents at 25$\mu$A per junction. At our target critical current density of 1kA/cm$^2$
the junction capacitance is about\cite{hypres_rules} 59 fF/$\mu$m$^2 \times 2.5\mu m^2 = 147$ fF. However, there is also the parallel capacitance of the modulation coil area between the junctions. We estimate there to be about $(50 \mu m)^2$ of such area. Taking the dielectric constant of SiO$_2$ to be 4.75 and the oxide thickness to be 0.1 $\mu m$ yields a parasitic capacitance of about 1 pF. Then the critical Stewart-McCumber \cite{stewart1968cvc,mccumber1968twl}  shunt resistance would be
$R_{shunt}=\sqrt{\Phi_0/2\pi I_0 C} = 3.6 \Omega$. Our design values for the shunt resistors were 4 Ohms and 2 Ohms. The resulting devices had resistances of about 2 Ohms. An earlier run, which had design values for the shunt resistors of 8 Ohms, resulted in hysteretic SQUIDs. These SQUIDs displayed relaxation oscillations\cite{adelerhov1995hsm} when operated in series with an array amplifier with high input inductance.\cite{huber2001dcs}  The contribution to the total SQUID flux noise $S_{\Phi}^{1/2}$ from thermal noise in the (2 Ohm)  shunt resistors in our susceptometers as designed is predicted to be\cite{tesche1977dcs} $S_{\Phi}^{1/2} \approx \sqrt{8 k_B T L^2/R} = 0.4 \times 10^{-6} \Phi_0/Hz^{1/2}$ at 4.2K.

\subsection{\label{sec:calculated PSF}Calculated magnetic response}
\subsubsection{Model}
 When the device dimensions are comparable to the London penetration depth (which we take to be 0.08 $\mu m$ in Nb) it is important to take into account the Meissner screening of the full 3-dimensional device geometry. For this purpose we followed a prescription given by Brandt,\cite{brandt2005tss} which we summarize here for completeness. The three-dimensional super-current density ${\vec j}$ in a magnetic field ${\vec H}$ is described by London's second equation:
\begin{equation}
\nabla \times {\vec j} = -{\vec H}/\lambda^2,
\label{eq:3dlondon}
\end{equation}
where $\lambda$ is the London penetration depth. For a film of thickness $d$ comparable or thinner than $\lambda$ in the $xy$ plane we integrate over $z$ to obtain
\begin{equation}
\nabla \times {\vec J} = -\vec{H}/\Lambda,
\label{eq:2dlondon}
\end{equation}
where ${\vec J}$ is the two-dimensional super-current density and $\Lambda \equiv \lambda^2/d$ is the Pearl length. Brandt defines a stream function $g(x,y)$ such that 
\begin{equation}
{\vec J}= -{\hat z} \times \nabla g = \hat{x} \frac{\partial g}{\partial y}-{\hat y}\frac{\partial g}{\partial x}.
\label{eq:stream}
\end{equation}
Then London's second equation becomes
\begin{equation}
H_z(x,y) = \Lambda \nabla ^2 g(x,y)
\label{eq:hz}
\end{equation}
The stream function $g(x,y)$ can be expressed as a density of current dipoles. Then the total $z$-component of the field in the plane of a 2-d superconductor is written as
\begin{equation}
H_z({\vec{r}}) = H_a({\vec{r}})+\int_S d^2{r'} Q({\vec r},{\vec{r}'})g(\vec{r}'),
\label{eq:hztotal}
\end{equation}
where $H_a({\vec{r}})$ is the externally applied field, and 
\begin{equation}
Q({\vec r},{\vec r}') = \lim_{z \rightarrow 0} \frac{2z^2-\rho^2}{4\pi(z^2+\rho^2)^{5/2}},
\label{eq:greens_function}
\end{equation}
with $\rho = |{\vec{r}-{\vec{r}'}}|$. Writing Eq. \ref{eq:hztotal} and \ref{eq:greens_function} as discrete sums:
\begin{equation}
H_z(r_i)=H_a(r_i)+\sum_j Q_{ij}w_jg(r_j),
\label{eq:discrete_hz}
\end{equation}
where $w_j$ is a weighting factor with the dimensions of an area, and
\begin{equation}
Q_{i\neq j} = \frac{-1}{4\pi |{\vec{r}}_i - {\vec{r}}_j|^3} \equiv -q_{ij}.
\label{eq:discrete_greens}
\end{equation}
$Q_{ij}$ is highly divergent for small values of $\rho$. Brandt notes that the total flux through the plane $z=0$ from any dipole source is zero in the absence of an externally applied field. Then for any ${\vec r}_i$ in the superconductor
\begin{equation}
0 = \int d^2 r' \, Q({\vec r}_i-{\vec r}') = \sum_j Q_{ij} w_j + \int _{\bar{S}} d^2 r' Q({\vec r}_i-{\vec r}').
\label{eq:zerosum}
\end{equation}
The discrete sum in Eq. \ref{eq:zerosum} is over the area inside the superconductor and the integral is over the area ($\bar{S}$) outside the superconductor. But the integral can be written as
\begin{equation}
\int_{\bar S} d^2 r' \, Q({\vec r}_i-{\vec r}') = \int_{\bar S} d^2 r' \frac{-1}{4 \pi | {\vec r}_i - {\vec r}' |^3} \equiv -C({\vec r}_i) = \oint \frac{d\phi}{4\pi R_i(\phi)},
\label{eq:circleint}
\end{equation}
where the last integral is over the angle $\phi$ between a fixed axis and a vector between the point ${\vec r}_i$ and a point on the periphery, and $R_i(\phi)$ is the length of this vector.  Returning to discrete sum notation,
\begin{equation}
Q_{ij} = (\delta_{ij}-1)q_{ij}+\delta_{ij}(\sum_{l \neq i}q_{il}w_l+C_i)/w_j
\label{eq:Qij}
\end{equation}
Eliminating $H_z$ from equation's \ref{eq:hz} and \ref{eq:discrete_hz} results in
\begin{equation}
H_a(r_i) = - \sum_j (Q_{ij} w_j- \Lambda \nabla^2_{ij})g(r_j)
\label{eq:ha}
\end{equation}
Inverting Eq. \ref{eq:ha} results in the solution for the stream function:
\begin{equation}
g({\vec r_i}) = -\sum_j K^{\Lambda}_{ij} H_a({\vec r_j}),
\label{eq:gfinal}
\end{equation}
with
\begin{equation}
K^{\Lambda}_{ij} = (Q_{ij} w_j - \Lambda \nabla ^2_{ij})^{-1}
\label{eq:Kfinal}
\end{equation}
We first calculate the $K^{\Lambda}_{ij}$ matrix given the geometry and Pearl length $\Lambda$ from Eq.'s \ref{eq:circleint}, \ref{eq:Qij}, and \ref{eq:Kfinal} (in that order), calculate the stream function from Eq. \ref{eq:gfinal}, and then calculate the total field anywhere in the same plane for a given source field from Eq. \ref{eq:discrete_hz}. The three components of the field for any position with $z_i \neq z_j$ are given by 
\begin{eqnarray}
\label{eq:stream_to_fields}
H_z(r_i)&=&H_a(r_i)+\sum_j w_jg_{j}\frac{2(z_i-z_j)^2-\rho^2}{4\pi((z_i-z_j)^2+\rho^2)^{5/2}} \nonumber \\
H_x(r_i)&=& \sum_j 3w_jg_j\frac{(z_i-z_j)(x_i-x_j)}{4\pi((z_i-z_j)^2+\rho^2)^{5/2}} \\
H_y(r_i)&=& \sum_j 3w_jg_j\frac{(z_i-z_j)(y_i-y_j)}{4\pi((z_i-z_j)^2+\rho^2)^{5/2}}. \nonumber
\end{eqnarray}

Following Brandt we replace a detailed (and time consuming) calculation of $C_i$ from Eq. \ref{eq:circleint} with the analytical expression for a rectangular area $|x| \le a$, $|y| \le b$ which encloses the  superconducting shapes of interest:
\begin{equation}
C(x,y)=\frac{1}{4\pi} \sum_{p,q} [(a-px)^{-2}+(b-qy)^{-2}]^{1/2}
\label{eq:rectangle_C}
\end{equation}
with $p,q = \pm 1$.

We used Delaunay triangulation \cite{delaunay1934sphere} to tile our surfaces, with a simplified version of the prescription by Bobenko and Springborn\cite{bobenko2006adl} to construct the Laplacian operator:
\begin{equation}
\nabla^2_{i,j} = \frac{1}{w} \sum_{j=1}^{N_i} (\delta_{i,j}-\delta_{i,i})
\label{eq:laplacian}
\end{equation}
where the sum is over the $N_i$  nearest neighbors of the $i^{th}$ vertex, and $w=ab/N_v$, with $ab$ the enclosing area (see Eq. \ref{eq:rectangle_C}) and $N_v$ the number of vertices in the triangulation. Eq. \ref{eq:laplacian} holds exactly for a square lattice, and also works well for a triangular lattice with sufficiently dense vertices.

Finally, Brandt provides a prescription for including externally applied currents. Assume for the moment that there is a delta function current $I$ at the inner edge of a superconducting shape with a hole in it. This is equivalent to applying an effective field
\begin{equation}
H^{\rm eff}_a = -I \sum_{\rm j\, in \, hole}(Q_{ij} w_j - \Lambda \nabla^2_{ij})
\label{eq:heff}
\end{equation}
The supercurrents generated in response to this field are described by the stream function
\begin{eqnarray}
g({\vec r}_i) &= - \sum_{\rm j \, in \, film} 
K^\Lambda_{ij} H^{\rm eff}_a ({\vec r}_j) \hspace{0.2in} &{\vec r_i }\, {\rm in \, film} \nonumber \\
&=I  & {\vec r_i }\, {\rm in \, hole} \nonumber \\
&=0 & {\vec r_i }\, {\rm outside \, film} 
\label{eq:geff}
\end{eqnarray}
The fields generated by the current are then calculated from Eq. \ref{eq:discrete_hz} as before.

Our devices consist of multiple levels of superconducting films. In our calculations we treated each film as 2-dimensional, with its in-plane (xy) shape given by our design files, but its z-position given by the average of the top and bottom heights of the film. In principle, the response of multiple films can be handled iteratively- using the sum of the responses of all of the films to the source field, and using this sum (plus the source) as the source for the next iteration. However, these calculations are quite time consuming, and the iterative technique does not converge quickly. In practice, to calculate the response to magnetic fields of our susceptometers we started with the superconducting film closest to the source, calculated its response to an externally applied field, used the source plus response field from this film as the source for the next film, etc. As we will see in Sec. \ref{sec:characterization}, these calculations, except for the smallest pickup loop, tend to overestimate the mutual inductance between the field coil and the pickup loop in our geometry by about 20\%. In our fits of the response of the susceptometers to various field sources, such an overestimation can result in fit heights that are lower than seems physically reasonable by a few tenths of a micron.

\subsubsection{Calculations of magnetometry}
\label{sec:mag_calcs}
\begin{figure}
\includegraphics[width=0.8\textwidth]{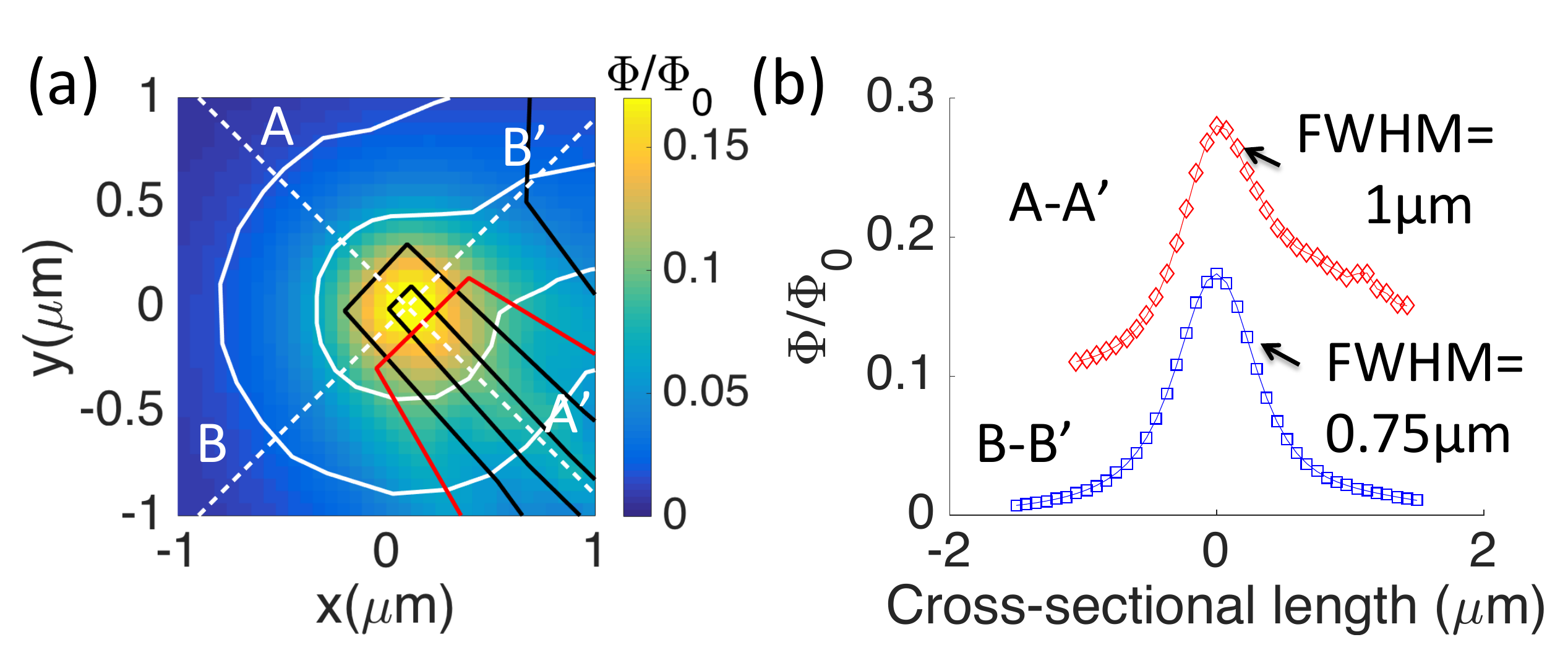}
\caption{
(a) Calculated flux signal (in units of $\Phi_0$) for a point monopole source with total flux $\Phi_0$ for a 0.2 $\mu$m inside diameter pickup loop susceptometer scanning in contact with the surface of the superconductor. The solid lines are outlines of the pickup loop/field coil layout. In this calculation the z-positions of these films are assumed to be the average heights of the levels as displayed in Fig. \ref{fig:three_level_process}. (b) The data displayed as diamonds and squares are cross-sections through the image in (a) as indicated by the dashed lines. Cross-section $A-A'$ is offset vertically by 0.1 $\Phi_0$ for clarity. 
}
\label{fig:Brandt_vortex}
\end{figure}

To calculate the response of our sensors to various field sources we substituted an assumed field distribution $H_a(\vec{r_j})$ into Eq. \ref{eq:gfinal} to find stream functions for the various superconducting shapes, used the first of Eq.s \ref{eq:stream_to_fields} to calculate the total field (source plus response fields) at the level of the pickup loop, and numerically integrated over the geometric mean area of the pickup loop to obtain the flux. 

An example is displayed in 
Figure \ref{fig:Brandt_vortex}, which shows the results of our calculations assuming a point source (monopole) vortex with total flux $\Phi_0$ (${H_a}({\vec{r}_j}) = \Phi_0 z_j/2\pi\mu_0 r_j^3$ in Eq. \ref{eq:gfinal}), with the zero in $z_j$ located at the top of the W2 layer - corresponding to the W2 shield in direct contact with the sample surface. The solid lines in this figure outline the various layers in the pickup loop/field coil region. One can see from Fig. \ref{fig:Brandt_vortex} (a) that the susceptometer response extends outside of the pickup loop region, due to flux spreading from the top of the W2 shield to the center of the pickup loop level. There are also ``tails'' to the field response due to flux penetration through and around the $W2$ shield in the area of the pickup loop leads. The diamond and square symbols in Fig. \ref{fig:Brandt_vortex} (b) are cross-sections along the dashed lines in Fig. \ref{fig:Brandt_vortex} (a). These cross-sections, which represent the ultimate spatial resolution of this sensor in the presence of a superconducting vortex, have full widths at half-maximum of 0.75 $\mu$m and 1 $\mu$m for cross-sections perpendicular and parallel to the leads respectively. 

\begin{figure}
\includegraphics[width=1.0\textwidth]{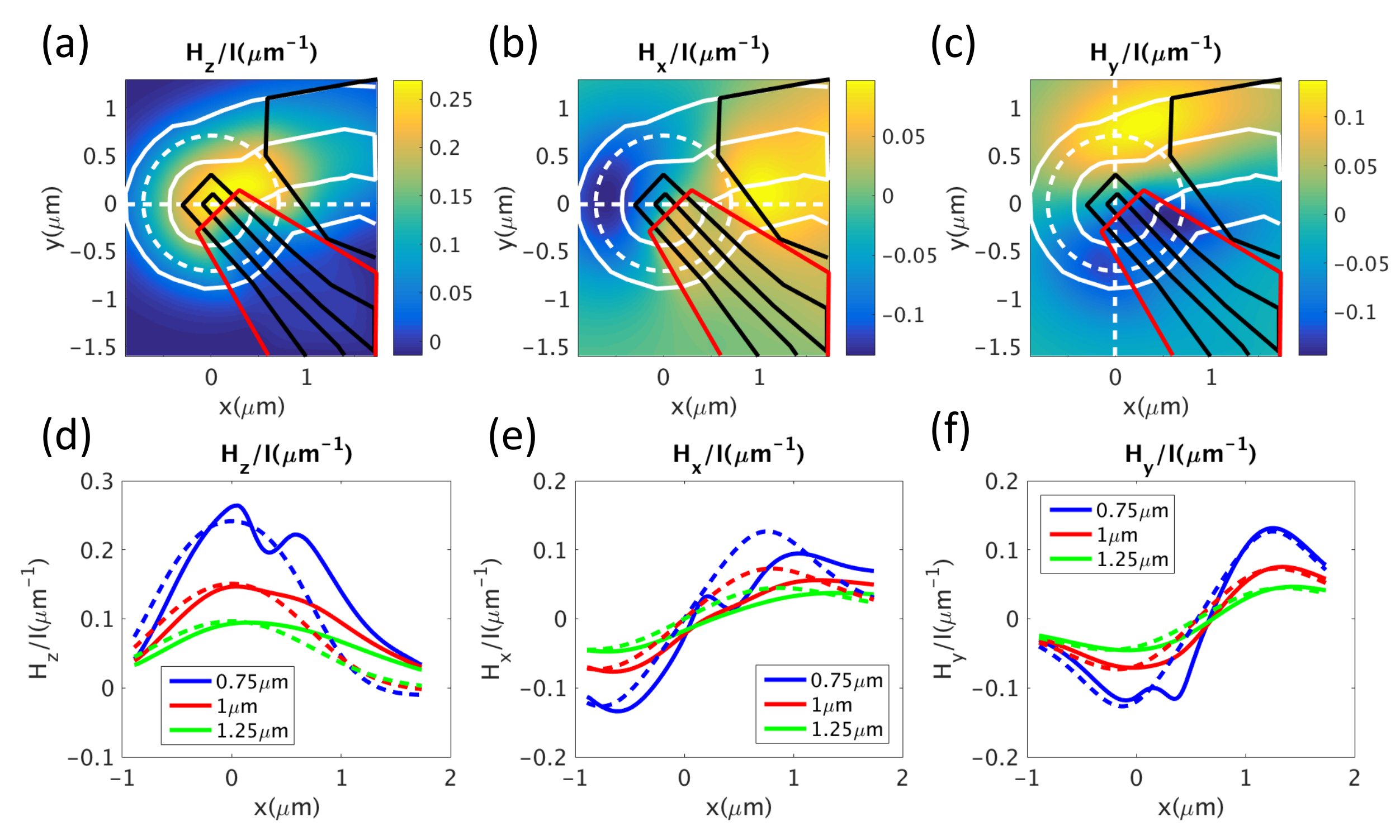}
\caption{Calculated components of the magnetic field in the $\hat{z}$ (a), $\hat{x}$ (b), and $\hat{y}$ (c) directions, normalized by the total current, due to a current in the field coil of our smallest pickup loop susceptometers, at a spacing of $z$=0.75 $\mu$m from the plane of the field coil. The solid lines overlaid on the field images are the layout of the pickup loop/field coil region. The solid lines in (d), (e), and (f) are cross-sections through the calculated fields at spacings of $z$=0.75, 1.0, and 1.25 $\mu$m along the positions indicated by the dashed lines in (a), (b), and (c) respectively. The dashed lines in (d), (e), and (f) are calculated fields from a circular, narrow wire, as indicated by the dashed circles in (a)-(c), for comparison.}
\label{fig:fc_fields}
\end{figure}

\subsubsection{Calculations of susceptibility}
\label{sec:susc_calcs}
To calculate susceptibility, we use Eq. \ref{eq:geff} to obtain the stream function of the field coil in the presence of an applied current. We then use the first of Eq.'s \ref{eq:stream_to_fields} to obtain effective fields to insert into Eq. \ref{eq:gfinal} for the other superconducting shapes. The magnetic fields at various spacings, calculated using Eq.s \ref{eq:stream_to_fields}, are displayed in
Figure \ref{fig:fc_fields} for our smallest pickup loop susceptometers.  One can see from the images of Fig. \ref{fig:fc_fields} (a)-(c) that the fields generated by the field coil are spread out by its finite width, and distorted by Meissner screening of the shield layers. The solid lines in Fig. \ref{fig:fc_fields}(d)-(f) are cross-sections along the straight dashed lines in Fig. \ref{fig:fc_fields}(a)-(c). For comparison, the dashed lines in Fig. \ref{fig:fc_fields}(d)-(f) represent the calculated fields from a circular, narrow wire of radius $c$ carrying a total current $I$:
\begin{eqnarray}
H_z &=& \frac{I}{4\pi} \int_0^{2\pi}d\theta \frac{c^2-c y \sin\theta - c x \cos\theta}{((x-c \cos\theta)^2+(y-c\sin\theta)^2+z^2)^{3/2}} \nonumber \\
H_x &=& \frac{I}{4\pi} \int_0^{2\pi} d\theta \frac{c z \cos\theta}{((x-c \cos\theta)^2+(y-c\sin\theta)^2+z^2)^{3/2}} \\
H_y &=& \frac{I}{4\pi} \int_0^{2\pi} d\theta \frac{c z \sin\theta}{((x-c \cos\theta)^2+(y-c\sin\theta)^2+z^2)^{3/2}}. \nonumber
\label{eq:thin_wire}
\end{eqnarray}
Here $c=0.79 \mu m$ is the geometric mean of the inside ($r_i=0.5 \mu m$) and outside ($r_o = 1 \mu m$) radii of the field coil: $c=\sqrt{(r_i^2+r_o^2)/2}$. The thin wire approximation is in reasonably good agreement with the full calculation, despite the distortions of the field caused by Meissner screening from the various superconducting films in our sensors.

The field coil fields can then be applied to a sample, the response fields calculated, and then integrated over the pickup loop to obtain a susceptibility. Results for a  superconducting shape using this procedure will be presented in Sec. \ref{sec:imaging tests}.
\section{\label{process}Fabrication}

\begin{figure}
\includegraphics[width=0.5\textwidth]{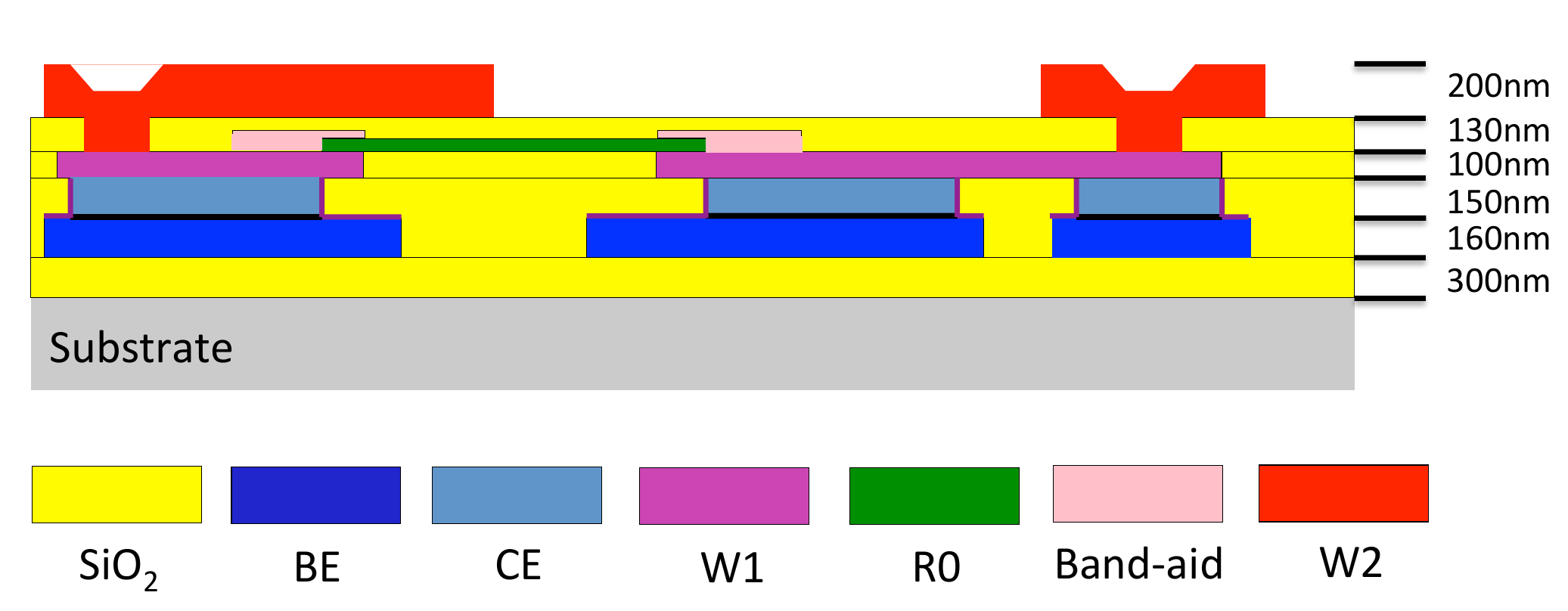}

\caption{Cross-section of the three level of metal, niobium trilayer process used. The yellow layers are insulating SiO$_2$, the green layer is a Au/Pd shunt resistor, and all other layers are superconducting Nb. Contacts to the trilayer through the Al$_2$O$_3$ insulator between $BE$ and $CE$ were used both as Josephson junctions (small areas) and vias (large areas).
}
\label{fig:three_level_process}
\end{figure}
We used a three level of metal, niobium trilayer\cite{gurvitch1983hqf} process on 200 mm diameter silicon wafers to fabricate these devices. This process is an extension of the original approach used to fabricate sub-$\mu$m Josephson junctions and SQUIDs using Nb-AlO$_x$-Nb trilayers in combination with chemical mechanical polish planarization.\cite{ketchen1991smup} An ASML 248 nm stepper was used for the lithography of all but the pickup loop features (in W1). The pickup loop was fabricated using an ASML 193 nm scanner, which had a minimum feature size and spacing of 0.2$\mu$m.  A cross-sectional view illustrating this process is shown in Figure \ref{fig:three_level_process}. All levels of metal except for the second wiring level were planarized by chemical-mechanical polishing (CMP). Briefly, the fabrication process was as follows: the silicon wafers were thermally oxidized to provide an insulating layer of 300 nm thick SiO$_2$. A Nb/Al$_2$O$_2$/Nb trilayer was deposited by sputter deposition. The aluminum was thermally oxidized to provide a 1kA/cm$^2$ critical current density for the Josephson junctions. The counter electrode of the trilayer was etched through the Al$_2$O$_3$ and approximately 50 nm into the Nb base electrode to form Josephson junctions and vias. A 40 nm thick layer of Nb$_2$O$_5$ was grown on the top surface of the trilayer, and then the base electrode was patterned by reactive ion etching. The pattern was filled with 150$^{\circ}$ C plasma-enhanced chemical vapor deposited (PECVD) silicon oxide and CMP'd back to the Josephson junction counter-electrode, removing the top layer of Nb$_2$O$_5$. The first wiring level of 100 nm of Nb was sputter deposited, patterned, filled with silicon oxide and planarized, a Au/Pd resistor layer (8 Ohms/square) was deposited and patterned by liftoff, a 100 nm thick 150$^o$C oxide was deposited and patterned for vias, and a 200 nm thick Nb second wiring layer was sputter deposited and patterned by RIE to complete the devices. Several iterations of design, fabrication, and test were performed. In the final run, which we report on here, there was poor conductance between the first wiring and shunt levels. This required remedial ``band-aid" structures of Nb to be added to make contact between the top of the first wiring level and the top of the shunt resistors. These band-aids provided extra topography, which tended to cause shorts through the intermediate insulating layer to the second wiring level, reducing our overall yield. Nevertheless, the process resulted in yields of ~50\% on the wafer that was most extensively tested (see Sec. \ref{sec:electrical}). Finally a 10 $\mu$m deep etch of the silicon wafers was performed, with care to keep within the thermal budget of the process. This allowed close alignment between the sample surface and the pickup loop without further processing of the SQUID sensors.

\section{\label{sec:characterization}Device characteristics}

\subsection{\label{sec:electrical}Electrical characteristics}
\begin{figure}
\includegraphics[width=0.5\textwidth]{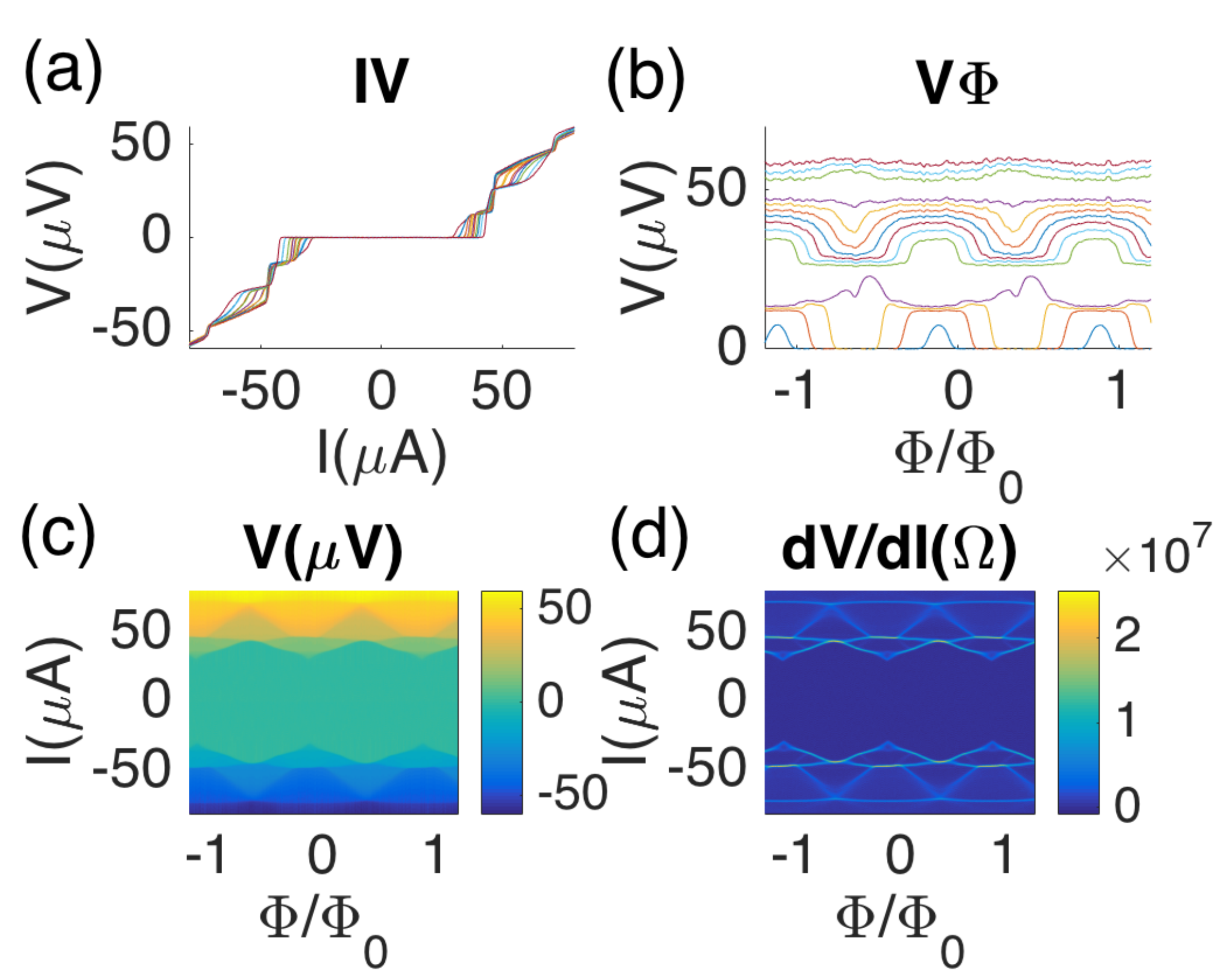}
\caption{Susceptometer electrical characteristics: (a) shows the current-voltage characteristics of a SQUID with an inner pickup loop diameter of 6 $\mu$m measured at 4.2 K for various fixed currents applied to the modulation coil, corresponding to an applied flux ranging from 0.5 $\Phi_{o}$ to 1.2 $\Phi_{o}$ in increments of 0.05 $\Phi_{o}$. 
(b) shows the voltage as a function of applied flux measured at various fixed applied currents through the SQUID.  The applied currents ranged from 33.3 $\mu$A to 80 $\mu$A in increments of 0.33 $\mu$A.
(c) is a mapping of current-voltage measurements in applied flux ranging from -1.2 $\Phi_{o}$ to 1.2 $\Phi_{o}$. The measurements were performed by measuring the voltage as a function of the current through the SQUID at various fixed values of the applied magnetic flux.  The applied flux ranged from -1.2 $\Phi_{o}$ to +1.2 $\Phi_{o}$.  The value of the voltage is indicated by the color scale, which ranges from -60 $\mu$V to 60 $\mu$V.
(d) shows the derivative with respect to the current, dV/dI , for the data shown in (c).
}
\label{fig:modulation_figure}
\end{figure}

\begin{figure}
\includegraphics[width=0.5\textwidth]{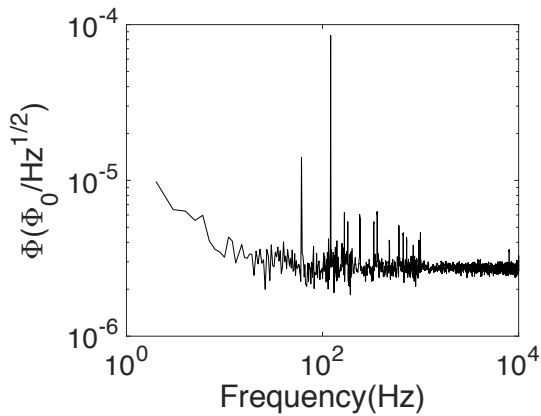}
\caption{Noise at 4 K for a 2 $\mu$m inside diameter SQUID susceptometer. The white noise floor is about 2.7 $\mu\Phi_0/Hz^{1/2}$, with a 1/f noise tail below 50 Hz. Our white noise floors varied between 2-4 $\mu\Phi_0/Hz^{1/2}$ depending on the particular device tested and bias conditions.
}
\label{fig:noise}
\end{figure}

Figure \ref{fig:modulation_figure} displays typical electrical characteristics of our susceptometers, measured at 4.2 K. Fig. \ref{fig:modulation_figure} (a) shows current-voltage characteristics for selected currents through the modulation coil. The modulation period is 0.15 mA, corresponding to a mutual inductance between the SQUID and modulation coil of 14 pH. The maximum in critical current was typically offset at random from zero flux, which we attribute to flux trapping. The maximum critical current of 45 $\mu$A is close to the design value of 50 $\mu$A (25$\mu$A/junction). Of the 150 SQUIDs characterized on one of the 7 wafers completed, 65 were good, as judged by a measurable critical current and modulation by both field coils, representing a yield of 43\%.  The critical currents of the susceptometers were independent of the pickup loop radii with a maximum value of 39.7 $\pm$ 8.5 $\mu$A. The modulation depths $(I_{c,max}-I_{c,min})/I_{c,max}$ were 0.51$\pm$0.01 for the 0.2, 0.6, and 2 $\mu$m pickup loop SQUIDs, and about 0.35$\pm0.03$ for the 6 $\mu$m SQUIDs. The average normal state resistances of the susceptometers were 1.1 $\pm$ 0.3 $\Omega$.
Table \ref{table:mutuals} lists the self-inductances for the SQUIDs, as a function of pickup loop inner diameter, inferred from the critical currents and modulation depths from standard SQUID theory,\cite{barone1982paj}. The inductance values for the three smallest pickup loop SQUIDs are comparable to our estimate of 50 pH from the design geometry, but the values for the 6 $\mu$m pickup loop SQUIDs are surprisingly high.  Two factors influence the contribution of the pickup loops to the total SQUID inductance. One is that the kinetic inductance becomes appreciable when the widths and thicknesses of the films making up the loops and leads become comparable to the penetration depth. The second factor is the length of unshielded leads. The 6 $\mu$m diameter pickup loop susceptometer has about 10 $\mu$m of unshielded leads, in comparison with the 2 $\mu$m susceptometer, which has less than 2 microns of unshielded leads. However, this latter factor cannot explain the anomalously high inductances inferred for the 6 $\mu$m susceptometers. The normal state resistances of our SQUIDs correspond to 2.2 Ohms/junction. Tesche and Clarke \cite{tesche1977dcs} estimate an optimized flux noise power floor from thermal sources of 
$S^0_{\Phi} \approx 8k_BTL^2/R$, which in our case suggests a flux noise of $(S^0_\Phi)^{1/2} \approx 0.4 \mu\Phi_0/Hz^{1/2}$ at 4.2K.
Our measured white noise floors (see Figure \ref{fig:noise}) are typically about a factor of 6 higher than this estimate. The best measured flux noises are typically 2-3 times higher than the Tesche-Clark estimate, so these devices display white noise somewhat higher than anticipated.

In addition to the SQUID modulation, there are also strong step-like structures in the current-voltage characteristics of our devices that we attribute to electromagnetic resonances driven by Josephson oscillations.\cite{fiske1964tmf} The combination of these resonances with standard SQUID interference produces the complicated, but continuous and non-hysteretic current-voltage characteristics seen in Fig. \ref{fig:modulation_figure}. Such resonances can be reduced by incorporating a damping resistor across the coaxial leads to the pickup loops.\cite{enpuku1986nco}

\begin{table*}[ht]
\caption{Pickup loop/field coil pairs}
\centering 
\begin{tabular}{c c c c c c c c c}
\hline
\hline                        
\multicolumn{2}{c}{Pickup loop} & \multicolumn{2}{c}{Field coil} & Shield &  Analytical & Numerical & Measured & Measured \\
d$_i(\mu$m) &  d$_o(\mu$m) & d$_i(\mu$m) & d$_o(\mu$m) & w($\mu$m) &  M ($\Phi_0$/A) & M($\Phi_0$/A) & M($\Phi_0$/A) & L(pH) \\
 [0.5ex]
\hline  
0.2 & 0.6 & 1 & 2   & 0.65 & 112.5 & 51 & 69$\pm$7  & 42$\pm$1  \\
0.6 & 1 & 2   & 3 & 0.77 & 173   & 188  & 166$\pm$4 & 43.2$\pm$1.6  \\
2   & 3 & 5 & 7 & 1.22 & 555   & 728  & 594$\pm$24 & 40.4$\pm$1.5 \\
6   & 7 & 12   & 16   & 1.56 & 1470  & -    & 1598$\pm$47 & 90.5$\pm$8 \\
[1ex] 
\hline 
\end{tabular}
\label{table:mutuals} 
\end{table*}

A good diagnostic test of our susceptometers is to measure the mutual inductances between the field coils and the pickup loops. For example, shorts between the ``band-aids" and the second wiring level are difficult to detect in IV measurements, but become immediately apparent in field coil/pickup loop mutual inductance measurements. Table \ref{table:mutuals} displays various parameters for the 4 pickup loop/field coil pairs reported in this paper. Here $d_i$ and $d_o$ are the inner and outer diameters of the pickup loop and field coil, and $w$ is the width of the upper shield where the leads connect to the pickup loop. The column labeled ``Analytical M" is the mutual inductance between the pickup loop and field coil calculated using the formula
\begin{equation}
M=\frac{\mu_0}{b\Phi_0}\left [ \pi a^2/4+w^2/3 \right ],
\label{eq:ketchen}
\end{equation}
where $a$ and $b$ are the effective diameters $d_{\rm eff} = \sqrt{(d_{\rm in}^2+d_{\rm out}^2)/2}$ of the pickup loop and field coil respectively. The second term in brackets in Eq. \ref{eq:ketchen} represents redirection of flux into the pickup loop from the shield. The analytical formula overestimates the redirection of flux into the pickup loop by the shield for the smallest pickup loop susceptometer. The ``Numerical M" values, calculated as described in Sec. \ref{sec:susc_calcs}, did not converge for the 6 $\mu$m inside diameter pickup loop susceptometers because there were too few vertices in the Delauney triangulation. The ``Analytical M'' calculations are in reasonable agreement with experiment except for the smallest pickup loop susceptometers, while the ``Numerical M'' calculations are high by about 20\% except for the smallest pickup loop susceptometers. The ``Measured L" column represents self-inductance derived using standard SQUID theory\cite{barone1982paj} from measurements of the critical current and modulation depth. 

The critical current of one of our 2$\mu$m inside radius susceptometers was 44 mA. This corresponds to a critical current density of 2.75$\times 10^7 A/cm^2$, in line with literature values\cite{huebener1975critical}, if we scale them down to our 1 $\mu$m linewidth for this field coil. There was no detectable change in the noise of this susceptometer as a function of field coil current up to the critical current. Since the peak z-component of the magnetic field produced by this field coil is 0.2T/A, the largest field that can be applied by the 2$\mu$m inside pickup loop radius susceptometer is about 9 mT. Our susceptometers are relatively insensitive to uniform magnetic fields: one of the 2 $\mu$m inside diameter pickup loop susceptometers had an effective pickup area to a uniform field perpendicular to the device plane of 5.3 $\mu m^2$, and no detectable change in its white noise floor, albeit an increase by a factor of 2 in the noise at 20 Hz, in perpendicular fields up to 1.4 mT. 
\subsection{\label{sec:imaging tests}Imaging tests}
\begin{figure}
\includegraphics[width=0.5\textwidth]{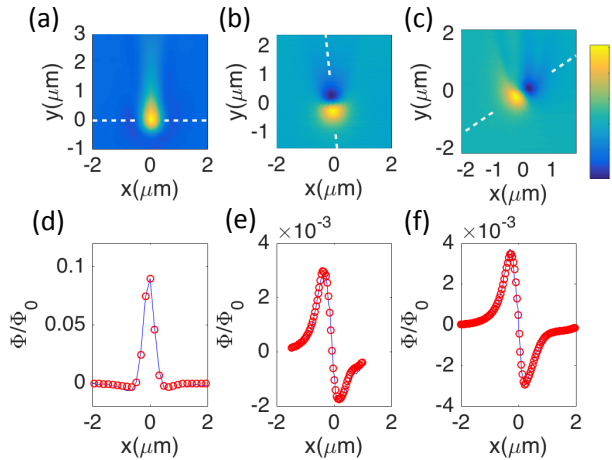}
\caption{Scanning SQUID magnetometry images of nanomagnets taken with a 0.2 $\mu$m inside diameter pickup loop susceptometer: (a) a nanomagnet with polarization perpendicular to the scan plane, (b) with polarization in plane (nearly) parallel to the pickup loop leads, and (c) with polarization in-plane at 56 degrees to the leads. The solid lines in (d)-(f) are cross-sections through the magnetometry images as indicated by the white dashed lines in (a)-(c). The red circles are numerical modeling as described in the text.
}
\label{fig:three_dipoles}
\end{figure}

We have tested the response of our 0.2 $\mu$m inside diameter pickup loop SQUID susceptometers to magnetic sources by imaging  nanomagnets, superconducting vortices and lines of current, and tested our susceptibility response using a superconductor.
\subsubsection{Magnetometry}
Figure \ref{fig:three_dipoles} displays scanning magnetometry images of 3 nanomagnets with different orientations of their magnetic moment. Parameters for these nanomagnets are given in Table \ref{table:nanomagnet}. Nanomagnet $a$ has the composition Ta(3)/Pt(5)/[Co(0.3)/Pt(1)]x5/Co(0.3)/Pt(4) (the numbers in parentheses are thicknesses in nm), and is magnetized primarily normal to the surface of the sample. Nanomagnets $b$ and $c$ are composed of CoFeB(4)/Pt(4) and are magnetized primarily in-plane. The solid lines in Fig. \ref{fig:three_dipoles} (d)-(f) are cross-sections through the images in Fig. \ref{fig:three_dipoles}(a)-(c) as indicated by the dashed white lines. 

The circles in Fig. \ref{fig:three_dipoles} (d)-(f) represent fits to the model described in Section \ref{sec:mag_calcs}, using 
\begin{equation}
H_a(\vec{r}_j)=m(3z_j(x_j \sin\theta\cos\phi+y_j\sin\theta\sin\phi+z_j\cos\theta)/r_j^5 -\cos\theta/r_j^3)/4\pi \hspace{0.3in} ({\rm dipole})
\end{equation}
in Eq. \ref{eq:gfinal} 
for a point dipole field source with moment $m$, with the fitting parameters $\theta$, $\phi$, $N=m/\mu_B$ the number of Bohr magnetons in the nanomagnet, and $h$ the spacing between the nanoparticle and the surface of the W2 shield. The fit values for $N$ (see Table  \ref{table:nanomagnet}) are significantly below the values calculated from the measured saturation magnetization of blanket films and the known volume of the nanomagnets , indicating that the moments are not completely aligned. We did not magnetize these samples in a high magnetic field before mounting them in the SQUID microscope. These nanomagnets did, however, enable us to demonstrate the spatial resolution of our susceptometers.

The nanomagnet in Fig. \ref{fig:three_dipoles}a generates a peak flux of 0.1 $\Phi_0$ through our 0.2 $\mu$m inside radius susceptometer, and has a fit value for $N=89\pm13\times 10^6 \mu_B$. Assuming a white noise floor of 2.7$\times10^{-6} \Phi_0/Hz^{1/2}$, this corresponds to a spin sensitivity of 2400 $\mu_B/Hz^{1/2}$. It has become traditional to express the spin sensitivity $S_n$ of small SQUIDs using a simple model proposed by Ketchen,\cite{ketchen1995dap} in which one calculates the flux from a point dipole source through a narrow wire loop with an effective size. For example, the spin sensitivity of a square superconducting loop of side $L$ becomes\cite{granata2016nano}
\begin{equation}
S_n^{1/2} =\frac{S_\Phi^{1/2}\pi L}{2\sqrt{2}\mu_0\mu_B},
\label{eq:spin_sensitivity}
\end{equation}
where $S_\Phi$ is the magnetic flux noise power spectral density. This would correspond to a spin noise of 170 $\mu_B/Hz^{1/2}$ for our flux noise and smallest pickup loop dimensions, if we use the geometric mean diameter for the characteristic pickup loop size. The discrepancy between these two estimates indicates that care should be taken when using simple formulas for estimating spin sensitivities.
\begin{table*}[ht]
\centering 
\begin{tabular}{| c | c | c | c | c | c | c | c |}
\hline
\hline                        
Nanomagnet &  $A$ &  $B$ & $t$ & $M_s$ &
Calc. $N$  &  Fit $N$  & Fit h  \\
 &  (nm) &  (nm) & (nm) & $10^6$(A/m) &
$10^6\mu_B$ &  $10^6 \mu_B$ & $\mu$m  \\
 [0.5ex]
\hline  
a & 250 & 250   & 6.5 & $ 0.65\pm0.03$ & $89\pm 13$ & $32+9-6 $ & -0.06+0.05-0.02    \\
b & 125 & 62  & 4 & $ 2.2\pm 0.1$ & $23\pm 3 $ & $5.0\pm 0.9 $ & 0.09+0.04-0.02   \\
c & 125 & 62   & 4 & $ 2.2\pm 0.1$ & $23\pm 3$ & $5.5\pm 1.3$ & 0.07+0.14-0.05    \\
[1ex] 
\hline 

\end{tabular}
\caption{Parameters of the nanomagnets imaged in Fig. \ref{fig:three_dipoles}. The first column corresponds to the labeling in Fig. \ref{fig:three_dipoles}.  $A$ and $B$ are the design semi-major and semi-minor axes respectively of the elliptically shaped nanoparticles, $t$ is the magnetic thickness and $M_s$ is the saturation magnetization at 4.2 K. The ``Calc. $N$" values are calculated from  $N =\pi A B t M_s/\mu_B$. The ``fit $N$" and $h$ values are from fits to the data as described in the text. Here $h$ is the spacing between the top surface of the susceptometer shield layer $W2$ and the nanomagnet, assumed to be a point dipole source. }
\label{table:nanomagnet} 
\end{table*}

\begin{figure}
\includegraphics[width=0.5\textwidth]{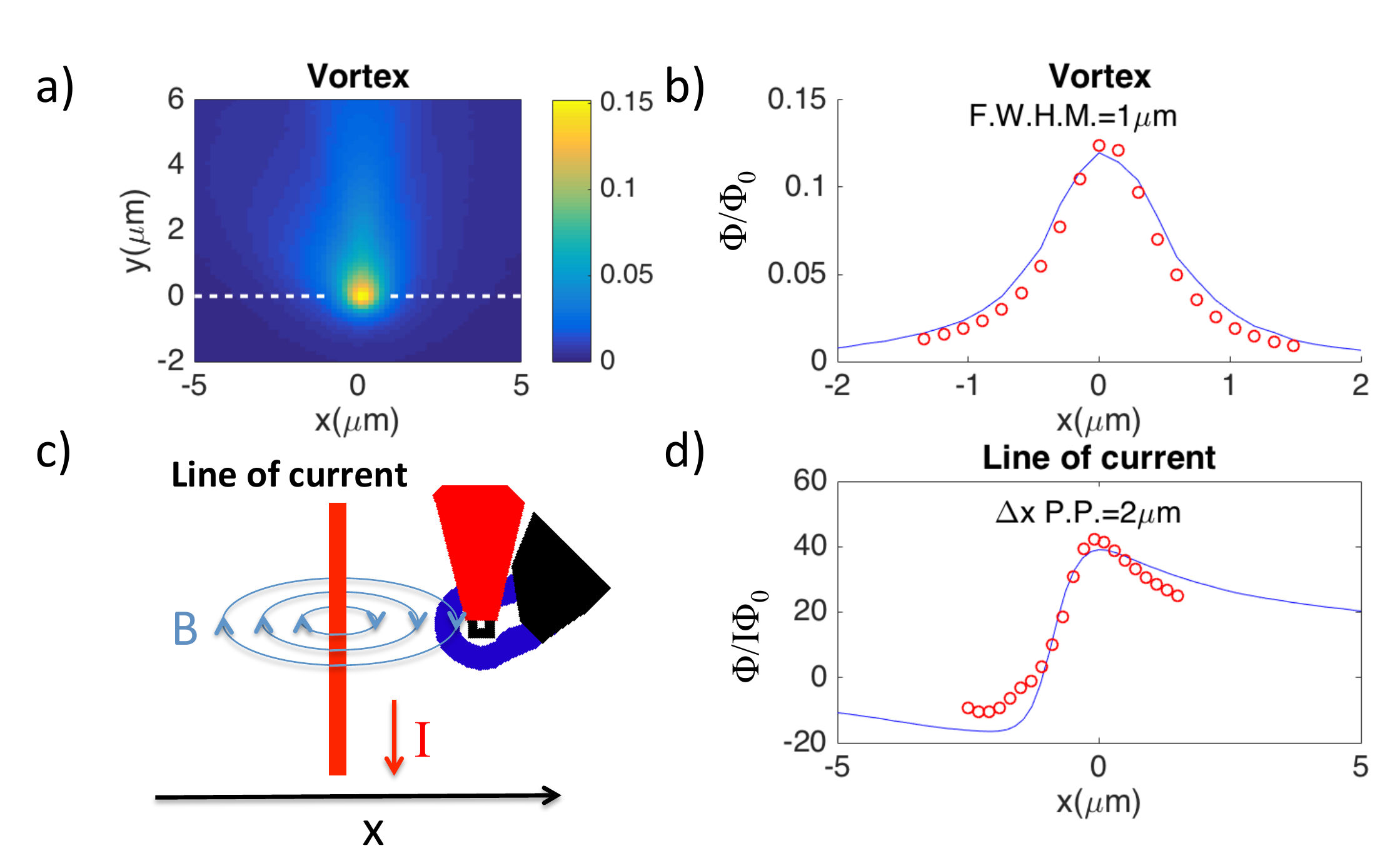}
\vspace{0.1in}
\caption{(a) Magnetometry image of a superconducting vortex. (b) The solid line is a cross-section along the dashed line in (a). The circles are a fit to the data. (c) Magnetometry signal from an 0.8 $\mu$m wide current carrying strip, with diagram showing the experimental geometry. (d) The solid line is the cross-section along the dashed line in (c).  The circles are the predictions of the model of Section \ref{sec:mag_calcs}.
}
\label{fig:two_sources}
\end{figure}

Figure \ref{fig:two_sources} (a) is a magnetometry image, with a separation between SQUID and sample of about 0.1 $\mu$m, of a superconducting vortex trapped in a 0.4 $\mu$m thick Nb film at 4.2K. The solid line in Fig. \ref{fig:two_sources} (b) is a cross-section along the dashed line in Fig. \ref{fig:two_sources} (a). The circles in Fig. \ref{fig:two_sources} (b) represent a fit of the model of Section \ref{sec:mag_calcs}, with 
\begin{equation}
H_a(\vec{r}_j)=\frac{\Phi_0z_j}{\mu_0r_j^3} \hspace{0.3in} ({\rm vortex})
\label{eq:vortex_ha}
\end{equation}
in Eq. \ref{eq:gfinal}
and with a single fit parameter $h=0.12\pm0.02\mu$m - the spacing between the top of the $W2$ shield layer and the surface of the sample. 

Fig. \ref{fig:two_sources} (c) is a cross-sectional image of a 0.8$\mu$m wide current-carrying Pt thin film wire composed of a 3 segment loop, with two long segments in the $x$-direction connected by a segment 200 $\mu$m long in the $y$-direction, imaged in contact. The solid line in Fig. \ref{fig:two_sources} (d) is a cross-section along the dashed line  in Fig. \ref{fig:two_sources} (c). The circles in Fig. \ref{fig:two_sources} (d) represent a calculation of Sec. \ref{sec:mag_calcs} for an infinitely narrow wire carrying current $I$ using the Biot-Savart law for the source field in Eq.\ref{eq:gfinal}:
\begin{equation}
H_a(\vec{r_j})=\frac{I}{4\pi}\int \frac{\hat{z}\cdot\vec{dl}\times\vec{r_j}}{|r_j]^3} \hspace{0.3in} {\rm line \, of \, current}
\label{eq:current}
\end{equation}
In this case it was assumed that the dipole and the $W2$ surface were in contact. 
The experimental line-width in this case is slightly broader than the calculation, perhaps because of the finite width of the current carrying wire.

As mentioned above, some of the spacings derived from our fits are smaller than is physically reasonable. For example, the fit value of $h$ for nanomagnet $a$ is negative: implying that the dipole source is inside the susceptometer shield.  Also, the fit heights for the vortex of Fig. \ref{fig:two_sources} are smaller than one would expect, given that vortex fields should spread at the surface of a superconductor as if the monopole source is a penetration depth $\lambda \approx 0.1\mu$m {\it below} the surface of the superconductor. Nevertheless, the good agreement between the experimental and calculated cross-section line-shapes gives us confidence that we have a nearly quantitative understanding of the magnetic field response of our susceptometers.
\subsubsection{Susceptibility}

\begin{figure}
\includegraphics[width=0.5\textwidth]{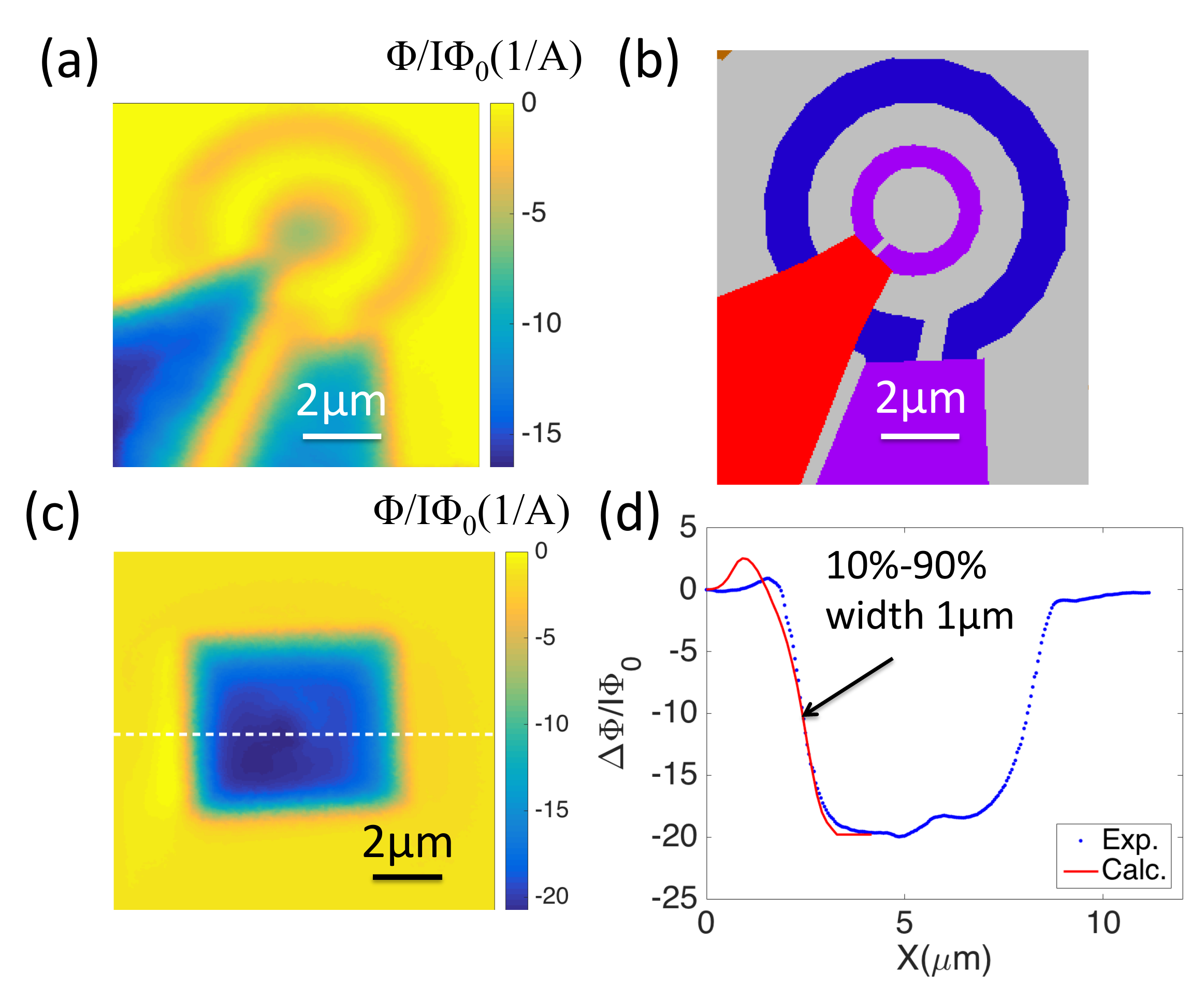}
\vspace{0.1in}
\caption{Susceptibility data taken with our 0.2$\mu$m inside diameter pickup loop susceptometer.  (a) Susceptibility image of the pickup loop/field coil region of a 2 $\mu$m inside diameter pickup loop susceptometer. (b) Layout for comparison with (a). (c) Susceptibility image of a 6 $\mu$m diameter square niobium pillar. (d) The dots are a cross-section along the dashed line in (c). The solid line are calculations as described in Sec. \ref{sec:susc_calcs}.
}
\label{fig:susceptibility_data}
\end{figure}

Our smallest pickup loop devices have higher spatial resolution for susceptibility as well as for magnetometry.  Examples are shown in Figure \ref{fig:susceptibility_data}. Figure \ref{fig:susceptibility_data}(a) displays susceptibility data of the pickup loop/field coil region of one of our 2 $\mu$m inside radius pickup loop devices taken with our smallest pickup loop SQUID, at 4.2K, with 1 mA current through the sensor field coil, and with the sensor scanning in contact with the sample. Figure \ref{fig:susceptibility_data}(b) displays the layout of the sample for comparison. The Nb film making up the pickup loop, which is 0.5 $\mu$m
 wide, is not quite resolved in this image. Figure \ref{fig:susceptibility_data}(c) shows a susceptibility image (taken under the same conditions as (a)) of a 6 $\mu$m wide square Nb ``pillar''. These pillars, composed of all of the layers in Fig. \ref{fig:three_level_process}, are about 750 nm high. They are used as fill to make the chip flat on average over a scale of a few tens of microns to assist in the CMP steps. The step in susceptibility indicated by the dots in Fig. \ref{fig:susceptibility_data}d has a 10\% to 90\% width of about 1 micron. The solid line in Fig. \ref{fig:susceptibility_data}d is modeling as outlined in Sec. \ref{sec:susc_calcs}, assuming a penetration depth for the pillar of 0.08 $\mu$m, and with the spacing $h$ between the pillar and the $W2$ surface of the susceptometer as a fitting parameter. The displayed best fit was for $h$ = 0.8 $\mu$m. This value is larger than seems reasonable from other measurements and the sample-sensor geometry. In addition, the calculated cross-section is slightly broader, and with a more pronounced overshoot, than the experiment. These discrepancies may be related to the finite height of the Nb pillar.  The full width of the 10\% to 90\% transition width was 1 $\mu$m wide, significantly narrower than the 2 $\mu$m full width at half maximum of the calculated field coil fields (see Fig. \ref{fig:fc_fields}), making it appear that the spatial resolution in susceptibility is determined primarily by the pickup loop size, as opposed to the field coil size.
 
\section{\label{sec:conclusions}Conclusions}
In conclusion, we have used a tri-layer niobium, fully planarized process with 0.2 $\mu$m linewidths and spacings, and a deep etch step, to fabricate scanning SQUID susceptometers with demonstrated sub-micron spatial resolution. These devices have pickup loops integrated into the body of the SQUID through well shielded, low inductance leads, integrated modulation coils for flux feedback, integrated field coils for spatially localized susceptibility measurements, and are fabricated using batch processing that produces tens of thousands of devices. 

\section{\label{sec:acknowledgements}Acknowledgements}
This work was supported by an NSF IMR-MIP Grant No. DMR-0957616. J.C.P. was supported by a Gabilan Stanford Graduate Fellowship and an NSF Graduate Research Fellowship, Grant No. DGE-114747. The nanomagnet sample fabrication at Cornell was supported by the NSF (DMR-1406333 and through the Cornell Center for Materials Research, DMR-1120296), and made use of the Cornell Nanoscale Facility which is supported by the NSF (ECCS-1542081). We would like to thank Micah J. Stoutimore for providing the Nb film used to study vortices.

\input{susceptometer_RSI.bbl}

\end{document}

%% file: susceptometer_RSI.bbl
%